\documentclass[prl,superscriptaddress,twocolumn,showpacs,amsmath,amssymb]{revtex4-1}
\usepackage{graphicx} 
\usepackage{epsfig}
\usepackage{dcolumn} 
\usepackage{color}
\usepackage{changes}

\begin{document}
\title{Tuning the ground state of cuprate high-critical-temperature superconducting thin films by nanofaceted substrates}
\author{G. Mirarchi}
\affiliation{Dipartimento di Fisica, Universit\`a di 
  Roma ``La Sapienza'', P.$^{le}$ Aldo Moro 5, I-00185 Roma, Italy}
\author{R. Arpaia}
\affiliation{Quantum Device Physics Laboratory, Department 
of Microtechnology and Nanoscience, Chalmers University of 
Technology, SE-41296 G\"oteborg, Sweden}
\author{E. Wahlberg}
\affiliation{Quantum Device Physics Laboratory, Department 
of Microtechnology and Nanoscience, Chalmers University of 
Technology, SE-41296 G\"oteborg, Sweden}
\author{T. Bauch}
\affiliation{Quantum Device Physics Laboratory, Department 
of Microtechnology and Nanoscience, Chalmers University of 
Technology, SE-41296 G\"oteborg, Sweden}
\author{A. Kalaboukhov}
\affiliation{Quantum Device Physics Laboratory, Department 
of Microtechnology and Nanoscience, Chalmers University of 
Technology, SE-41296 G\"oteborg, Sweden}
\author{S. Caprara}
\affiliation{Dipartimento di Fisica, Universit\`a di 
  Roma ``La Sapienza'', P.$^{le}$ Aldo Moro 5, I-00185 Roma, Italy}
\author{C. Di Castro}
\affiliation{Dipartimento di Fisica, Universit\`a di 
  Roma ``La Sapienza'', P.$^{le}$ Aldo Moro 5, I-00185 Roma, Italy}
\author{M. Grilli}
\affiliation{Dipartimento di Fisica, Universit\`a di 
  Roma ``La Sapienza'', P.$^{le}$ Aldo Moro 5, I-00185 Roma, Italy}
\author{F. Lombardi}
\affiliation{Quantum Device Physics Laboratory, Department 
of Microtechnology and Nanoscience, Chalmers University of 
Technology, SE-41296 G\"oteborg, Sweden}
\author{G. Seibold}
\affiliation{Institut f\"ur Physik, BTU Cottbus-Senftenberg, 
D-03013 Cottbus, Germany}
\begin{abstract}
Anisotropic transport properties have been assessed in a number 
of cuprate superconductors, providing evidence for a nematic state. 
Here, we analyze recent experimental data for ultrathin 
YBa$_2$Cu$_3$O$_{7-\delta}$ (YBCO) films, where nematicity is 
induced via strain engineering, leading to a suppression of 
charge density wave scattering along the orthorhombic $a$-axis 
and a concomitant enhancement of strange metal behavior along the 
$b$-axis. It is shown that the anisotropic properties strongly 
depend on the substrate, which we characterize by atomic force 
microscopy (AFM). Based on the AFM data, we provide a microscopic 
model that can account for the absence (presence) of nematicity 
and the resulting transport properties in films grown on SrTiO$_3$
(MgO) substrates.
\end{abstract}
\maketitle

The ground state of cuprate high-critical-temperature superconductors 
(HTS) hosts some of the most intricate quantum phases in solid 
state physics, where charge, spin and orbital orders are 
intertwined~\cite{keimer2015quantum, fradkin2015colloquium,kloss16,proust19}. In these materials,  
superconductivity emerges from a metallic state, named ``strange 
metal” \cite{phillips22}, with properties that cannot be accounted for by the 
Fermi liquid theory. In this state, the quasiparticle picture for 
electron excitations breaks down and a robust $T$-linear dependence 
of the resistivity is observed in transport, as if the details of 
the scattering events become irrelevant in a certain 
temperature range \cite{varma16,zaanen19,hussey21,hartnoll22,mirarchi22}.

Modeling the strange metal state has proved to be a task of 
paramount difficulty, because of the strong electron-electron
interactions. Recent work therefore has addressed the problem of $T$-linear resistivity within advanced computational methods applied to the two-dimensional Hubbard model~\cite{Tom2019,wu22}.
Other approaches, involving critical fluctuations in the vicinity of a quantum critical point \cite{cha20,dumi22,Christos2023}, e.g., based on short ranged dynamical charge density fluctuations \cite{arpaia19, arpaia2022signature}, were also able to reproduce the fingerprint of this state of matter~\cite{sei20,caprara22}. Holographic duality, which maps the 
physics of strongly interacting systems onto the classical theory 
of gravity in higher dimensions, is a further pathway to obtain 
the $T$-linear resistivity \cite{Zaanen2022, baggioli2023}.

Though encouraging, all these theoretical attempts need further developments; they have to be predictive for the  physical mechanisms setting in when the $T$-linear resistivity breaks down.   The tuning of the ground state of HTS beyond oxygen doping, e.g., as recently demonstrated by various mechanical strain experiments~\cite{mito2017, kim2018, kim2021, boyle2021, choi2022, barber2022, nakata2022, guguchia2023, gupta2023}, can give additional information about the strange metal phase, its origin and the intertwining with the various local orders, including charge density waves (CDW)~\cite{tran95,abba05,GhiringhelliCDW,chang12,comin16, arpaia2021charge}. These new inputs, coming from the experiments, could be fed into theories to discriminate about the various scenarios.

Recent experiments on nm-thick YBCO films~\cite{eric20}  have demonstrated that the strain induced by the substrate strongly modifies the ground state of the material  affecting the Fermi surface topology,  the strange metal phase and the CDW order.   In ultrathin films (thickness $t$ = 10 nm, i.e., about 8 unit cells)  of YBCO grown on MgO (110) substrates,  one observes a strongly nematic Fermi surface which  cannot be accounted for  by the anisotropy induced by the CuO chains.  In addition, the CDW order,  that is biaxial in bulk systems and thick films, becomes unidirectional along the $b$-axis direction.  But, more importantly, the strange metal phase does not break down at the pseudogap temperature, as it happens in the underdoped regime, 
rather it extends  down to much lower temperatures.

It is natural to ask what the role of strain is in nm-thick 
films. Is it just the strain-induced geometrical modification 
of the unit cell that is important or the substrate-film 
interface coupling induces more subtle effects, 
with possible interplay with the intertwined orders? 
To answer these questions, we have compared the resistivity 
measurements of 10\,nm thick YBCO films grown on two substrates, 
MgO (110) and SrTiO$_3$ (001) with different lattice parameters 
and surface morphology. We show that the deviations from a 
bulk-like behavior are strongly dependent on these parameters 
and are observed only on MgO (110). Based on the morphological 
analysis of the substrates, we have  developed a model which 
describes different film-substrate coupling, depending on 
the extension and morphology of the surface facets. The results 
can nicely reproduce the nematic Fermi surface in the case of 
MgO, while confirming an unaltered ground state with the 
SrTiO$_3$ (STO) substrate. Starting from a nematic Fermi surface we 
can retrive the unidirectional CDW observed in the experiment 
and discuss the possible implication on the strange metal state. 

\begin{figure}[htbp]
\includegraphics[angle=-0,scale=1]{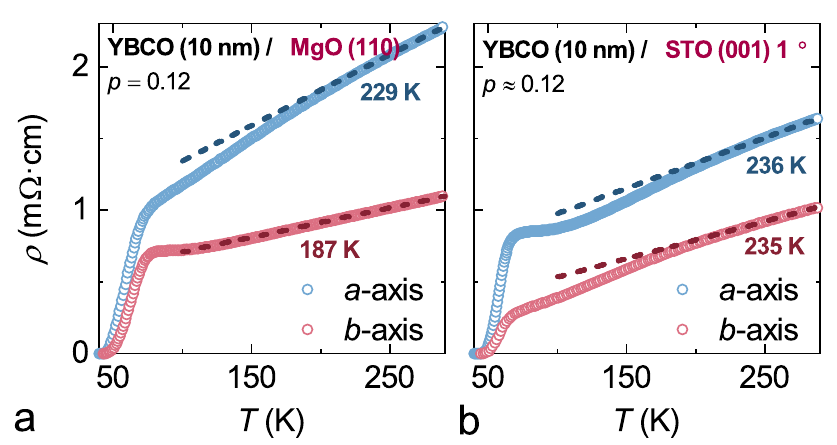}
\caption{Strain dependence of the in-plane resistivity of underdoped 
($p \approx 0.12$) ultrathin ($t = 10$ nm) YBCO thin films.
(a) The resistivity versus temperature $\rho$($T$) along the $a$- and 
$b$-axis of a film grown on a (110) oriented MgO substrate. 
The dashed lines are the linear fits of the curves for $T > 260$\,K. 
Close to these lines, the temperatures below which the resistivity deviates by $1\%$ from the linear behaviour are reported. (b) Same as panel (a), but for a film grown on a 
$1^\circ$ vicinal angle (001) oriented STO substrate.}
\label{RT}
\end{figure}

Figure \ref{RT}(a) shows the temperature dependence of 
the resistivity $\rho$, measured in two devices oriented along 
the YBCO $a$- and $b$-axis, realized in an underdoped ($p=0.12$) 
and untwinned 10\,\,nm thick film grown on a MgO substrate. 
The resistivity anisotropy ratio at $T=290$\,K, defined by 
$\rho_\text{a}$(290\,K)/$\rho_\text{b}$(290\,K) is larger than 
2, a value much higher than what expected for 
this level of doping. The slopes of $\rho_\text{a}(T)$ 
and $\rho_\text{b}(T)$ are quite different, which can be 
attributed to a different Fermi velocity along $a$- and $b$-axis, 
as extensively discussed in Ref.~\cite{eric20}. This hints at
a nematic Fermi surface. In these samples, both the strange
metal state and the CDW become very anisotropic. The 
$T$-linear behavior along the $b$-axis is extended down to 
$T_\text{L} = 187$\,K, while along $a$-axis $T_\text{L}$ is the 
same as in bulk samples. At the same time, the CDW measured 
by resonant inelastic X-ray scattering (RIXS) becomes 
unidirectional, along the $b$-axis~\cite{eric20}. 

In underdoped, untwinned, 10\,\,nm thick YBCO films grown on 
(001) oriented STO substrates with a vicinal angle, the 
transport properties along $a$- and $b$-axis are more conventional 
and similar to those observed in bulk samples. Here, in an
underdoped ($p=0.12$) film the slopes of the 
$T$-linear resistivity are very similar for $\rho_\text{a}(T)$ 
and $\rho_\text{b}(T)$ [see Fig.\,\ref{RT}(b)]. The Fermi 
velocities along $a$- and $b$-axis are therefore comparable, 
which implies a rather isotropic Fermi surface. In addition, 
the strange metal state and the CDW do not show any anisotropy, 
differently from the MgO case. Both $\rho_\text{a}$ and 
$\rho_\text{b}$ deviate from linearity at the same temperature 
and the value of $T_\text{L}$ agrees with previous measurements 
on bulk samples. The CDW measured with RIXS remains bidirectional. 

The large difference of the transport properties of films grown on 
MgO and on STO, despite the identical level of doping, suggests 
that the nematicity, which is conspicuous only in films grown on 
MgO, cannot be explained taking merely into consideration the CuO 
chains along the $b$-axis, since these are present and equally filled in both systems. 
To shed light on this peculiar anisotropic state, we focus on 
the structure of the substrate surface and on the interface it makes with the YBCO films.

\begin{figure}[htbp]
\includegraphics[angle=-0,scale=0.25]{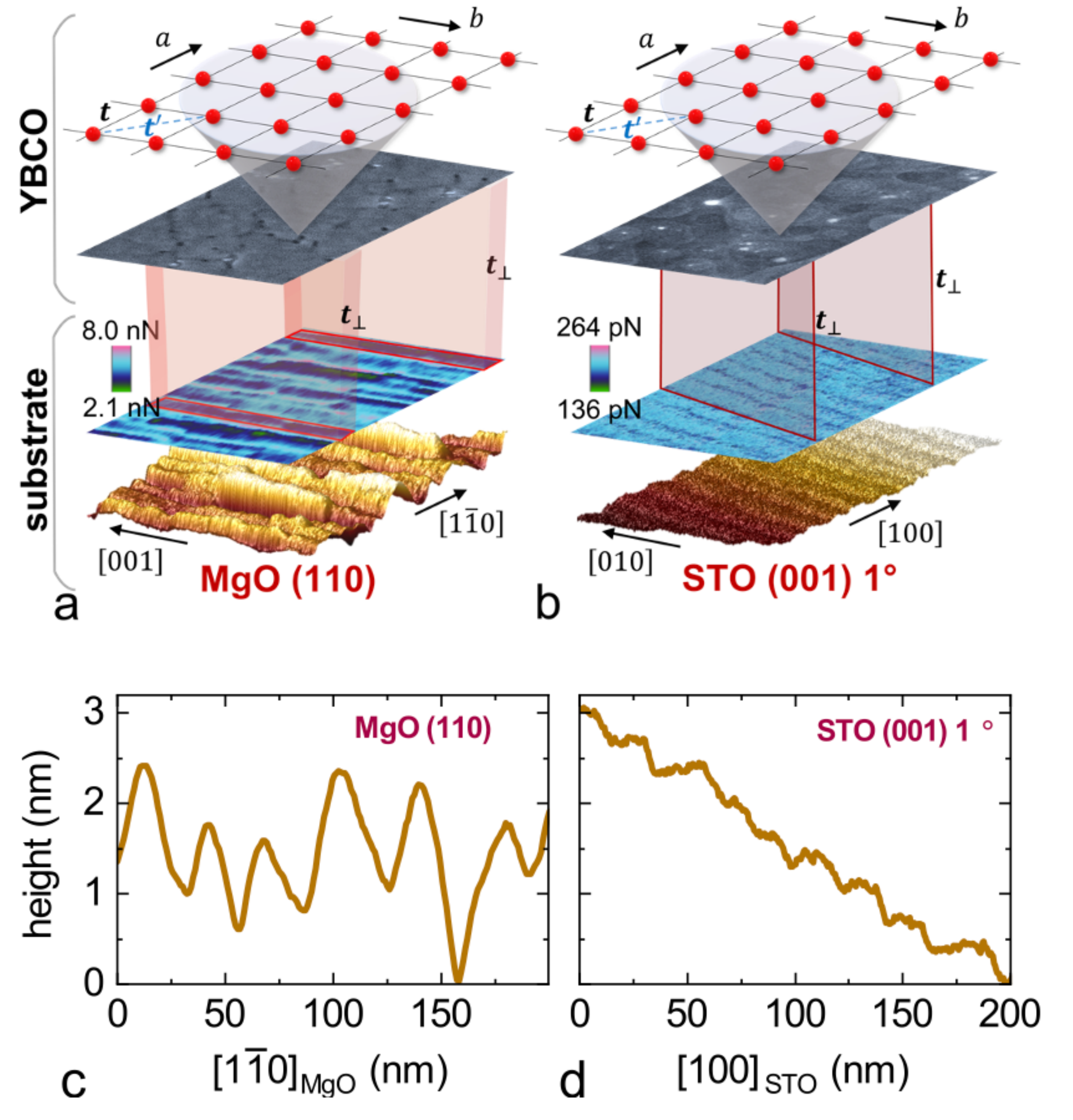}
\caption{Analysis of the substrate surface and modelization of 
the interface with the YBCO film. (a) From the bottom: TM AFM 
image of a 250$\times$250\,nm$^2$ region of the annealed MgO (110) 
substrate (yellow-brown map). 
The chemical contrast caused by the elongated and 
rather high nanofacets, running along the [001] MgO direction, 
is detected in the very same area by PFQNM AFM adhesion image (bluish map). 
The difference in coordination between the atoms in the valleys 
between the facets and those along the facet edges is large and 
gives a strong chemical contrast, involving large areas 
of the region under investigation. The YBCO film grown on 
top is represented by a 
scanning electron microscopy image. The planar tight-binding structure of the YBCO planes is the topmost 
enlargement at the interatomic scale (red dots) with the nearest neighbor hopping parameter $t$ and the next nearest neighbor hopping parameter $t'$. These atoms are those coupled to the under-coordinated regions of the substrate by a coupling parameter $t_\perp$ (see the wide red slabs). (b) Same as panel (a), but for 
a YBCO film grown on a $1^\circ$ vicinal angle (001) oriented 
STO substrate. Here, the difference is given by the areas with 
under-coordinated atoms, on the edges between adjacent 
atomic-scale-high steps, which are narrow and give a 
very small chemical contrast: the net effect is a tiny coupling 
between film and substrate. (c) TM AFM linescan along 
the [$1\bar{1}0$] MgO direction. (d) Same as panel (c), but 
along the [100] STO direction. Here, the height of the steps is 
much smaller than for the facets on the MgO surface, resulting 
in a film much less embedded in the substrate matrix.
}
\label{model}
\end{figure}

The surface of (110) oriented MgO substrates is characterized 
by elongated nanofacets, running along the [001] direction. They 
are the consequence of a surface reconstruction, obtained 
by annealing the substrate in oxygen atmosphere, which is 
instrumental to achieve untwinned YBCO films~\cite{arpaia192}.
As demonstrated by the tapping mode (TM) atomic force microscopy 
(AFM) investigation [see Figs.\,\ref{model}(a-c)], these 
nanofacets have an average height of about 1.5\,\,nm and a nearly triangular profile [see Fig. \ref{model}(c)].
This results in a very anisotropic atomic lattice, where large 
areas along the facet edges consist of under-coordinated atoms. 
This picture is confirmed by peak force quantitative nanomechanics 
(PFQNM) AFM adhesion analysis [see Fig.\,\ref{model}(a)], where 
the chemical contrast at the surface of the MgO substrate can be 
investigated. The difference in coordination between the atoms 
in the valleys and atoms along the facet edges is very large, 
and affects a substantial area of the substrate surface. When YBCO 
is deposited on top of this surface [see the scanning electron 
microscopy image in Fig.\,\ref{model}(a)], it is subject to a 
very anisotropic strain, resulting in an untwinned film, with the 
$a$-axis and the $b$-axis respectively aligned perpendicular and 
parallel to the MgO facets~\cite{arpaia192}. At the interface, 
the film is embedded in this articulated and anisotropic 
substrate matrix, with its unit cell height comparable to 
the nanofacet height. It is then natural to expect a 
strong coupling between the deposited film and the substrate, 
mainly occurring in the areas where the substrate atoms are 
under-coordinated. This strong coupling is based on the 
hybridization between the two layers to saturate these bonds.

We now face the problem of modelling this heterogeneous structure with 
disorder on the scale of tens of nanometers to get the electronic structure of YBCO. This model is based on interactions on atomic (i.e., over 0.3-0.4 nm) scales.
In Figure \ref{model} the YBCO layer is represented by the red-dot lattice, with 
a tight-binding electronic structure with the nearest 
neighbor hopping parameter $t$ and the next nearest 
neighbor hopping parameter $t'$. To model the more or 
less strong chemical coupling of the CuO$_2$ plane and the MgO or STO substrates, we introduce 
a hopping $t_\perp$ to the elongated wide 
regions of the substrate having under-coordinated atoms. 
According to the elongated shape of the facets in MgO 
(or of the shallower steps in STO, see below), 
only elongated (actually one-dimensional) regions of the CuO$_2$ lattice
running along the $b$-axis are endowed by the additional hopping $t_\perp$.
The virtual hopping of charge carriers to the substrate atoms 
induces an effective repulsion between the atomic levels
of film and substrate and, by eliminating the substrate
degrees of freedom, leads to an effective potential
$V_{\text{eff}}=\frac{1}{2}(\sqrt{V_{\text{sub}}^2+4t_\perp^2}-V_{\text{sub}})$
for the coupled atoms of the film, where $V_{\text{sub}}$ 
is the on-site energy of the coupled substrate atoms. 
The one-dimensional strips of atoms with the added $t_\perp$ hopping 
(in turn resulting in the above effective potential) are randomly distributed 
on the lattice, and therefore they can then be treated by a standard coherent potential approximation (CPA) 
(see Supplemental Material~\cite{SM}).

Having assessed the YBCO/MgO system, one might ask which picture 
can better describe the YBCO/STO system. Here, the scenario is 
in principle very similar. The $1^\circ$ vicinality of the 
substrate, instrumental to achieve untwinned YBCO films, results 
in step edges, having an average height of 0.39\,\AA~[the lattice 
parameter of STO, see Figs.\,\ref{model}(b-d)]. These steps induce 
an anisotropic strain, favoring the growth of untwinned YBCO 
films with the $b$-axis ($a$-axis) oriented parallel (perpendicular) 
to the elongated steps, which is the same epitaxial relation 
occurring on MgO substrates. An effective potential 
$V_\text{eff}$ can also be used in this case to describe the 
coupling between film and substrate. However, the coupling here 
is much weaker. The reason stems from the  atomic-scale-high steps 
characterizing the STO substrates, giving rise to a tiny chemical 
contrast [see Fig. \ref{model}(b)], involving much narrower areas 
of the surface of the substrate (the step edges are much sharper 
and reduced in height compared to the smooth slopes of the MgO 
nanofacets). In our model, the main difference between the 
two substrates is therefore given by the substantial different 
values of $V_\text{eff}$: $|V_\text{eff}|\sim t$, i.e., is of the 
order of the in-plane hopping $t$ in case of MgO, playing a 
significant role in building up the YBCO electronic structure; for 
the STO substrate, instead, $|V_\text{eff}|\ll t$.

We have now all the information to build the hamiltonian of our system where the interface coupling between YBCO and substrate plays a crucial role. As noticed above, the steps and facets of the substrates occur on the scale of some tens of nanometers,
while the atomic model for the electronic structure considers atomic scale of the order
of unit cells. This is also the scale on which the random distribution of the 
$b$-oriented strips feeling $V_\text{eff}$ is build in the tight-binding model which reads
\begin{equation}\label{eq:ham}
H= \sum_{ij,\sigma}t_{ij}c_{i,\sigma}^\dagger c_{j,\sigma}
+V_\text{eff} \sum_{n}\sum_{iy=1}^L c_{{\bf R}_n+iy {\bf b}}^\dagger c_{{\bf R}_n+iy {\bf b}}\,.
\end{equation}
Here, $t_{ij}$ includes hopping processes between nearest ($\sim t$) and next-nearest
($\sim t'$) neighbors, $V_\text{eff}$ is the effective potential along the $b$-oriented strips
with length $L$ starting at a random site ${\bf R}_n$. Due to the large anisotropy of both nanofacets and steps we further set $L\to\infty$
and we encode the fraction of regions affected by the nanofacets or steps (corresponding to the red slabs in Fig. \ref{model}) by the parameter $f$. In the model Eq. (\ref{eq:ham}) $f$ then corresponds to the fraction of sites on which the potential $V_\text{eff}$ acts. 

\begin{figure}[htbp]
\includegraphics[angle=-0,scale=1]{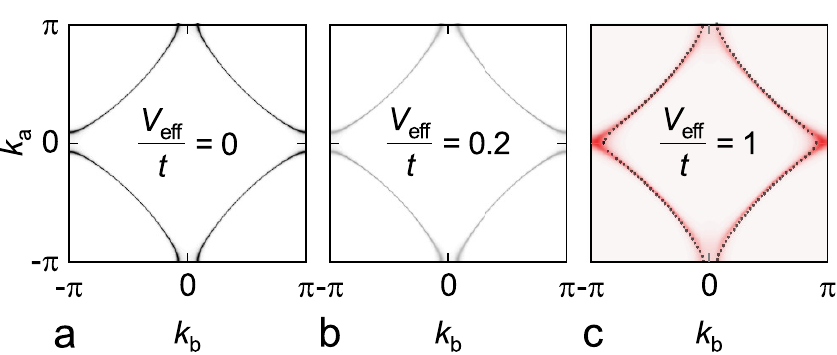}
\caption{Fermi surface within the CPA for different values of 
the effective potential $V_\text{eff}/t$. (a) $V_\text{eff}/t = 0$; 
(b) $V_\text{eff}/t = 0.2$; (c) $V_\text{eff}/t = 1$. In all 
cases, the concentration of one-dimensional strips is 
$f=0.15$, the next-nearest neighbor hopping is $t'/t=-0.15$ and 
the hole concentration is $p=0.12$. The Fermi surface obtained 
in panel (c) resembles that inferred in Ref.~\cite{eric20} to 
describe the transport properties of ultrathin YBCO films on 
MgO substrates. The black dotted line corresponds to a 
tight-binding parametrization derived from $t_\text{b,a} = 
t(1 \pm \alpha)$ and anisotropy parameter $\alpha = 0.015$.
}
\label{fscpa}
\end{figure}

Owing to the random character of the stripes and since we expect a smooth (on atomic scales) variation of the effective potential, we use a standard CPA
(see ~\cite{SM} for additional 
details). This allows to compute the self-energy and hence 
determine the Fermi surface for this disordered system. 
Figure\,\ref{fscpa} shows the Fermi surfaces that result upon 
varying the effective potential while keeping the parameter $f=0.15$ constant.  
For small $V_\text{eff}/t=0.2$
the Fermi surface [see Fig.\,\ref{fscpa}(b)] does not differ from 
the unperturbed one [see Fig.\,\ref{fscpa}(a)] and preserves the
$C_4$ symmetry. Instead, when the effective potential 
becomes significant, i.e., of the order of the in-plane hopping 
$t$, which is the case for films grown on MgO, the Fermi surface 
becomes nematic, i.e., anisotropic and distorted in the regions
around the points $(0,\pi)$ and $(\pi, 0)$ of the
first Brillouin zone [see 
Fig.\,\ref{fscpa}(c)], resembling the Fermi surface 
inferred in Ref.~\cite{eric20} to explain the in-plane 
transport anisotropy observed in ultrathin YBCO films on MgO 
[see Fig.\,\ref{RT}(a)]. 

This nematic Fermi surface has strong implications for the 
symmetry of the resulting charge collective excitations, which 
can affect the anisotropic transport properties of the films. 
In the following, we will demonstrate that the Fermi surface 
in Fig.\,\ref{fscpa}(c), resulting from the anisotropy imposed 
by the MgO nanofacets, gives rise to a unidirectional CDW along the 
$b$-axis, in agreement with what we have observed in underdoped 
($p = 0.12$) 10\,nm thick YBCO films grown on MgO~\cite{eric20}.
To this purpose we adopt the framework of frustrated phase separation, where 
the charge instability is not driven by nesting properties of the Fermi
surface, but merely results from the electron-electron Coulombic repulsion 
frustrating the phase separation induced by a variety of attractive 
mechanisms on a strongly correlated electron 
system~\cite{emerykivelson93,raimondi93,cast95,becca,andergassen}.

\begin{figure}[htbp]
\includegraphics[angle=-0,scale=1]{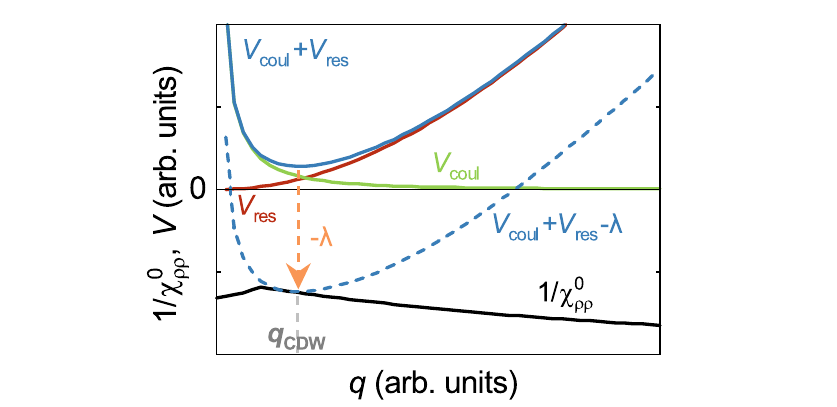}
\caption{Schematic representation of the CDW instability within the frustrated phase separation 
scenario.  The effective random-phase-approximation (RPA) 
expression $1/\chi_{\rho\rho}^{0}(q)-V_\text{tot}(q)=0$ where 
$\chi_{\rho\rho}^0(q)$ is the charge susceptibility of the 
non-interacting quasiparticle system. 
The total interaction $V_\text{tot}(q)$ 
(blue dashed line) is given by the sum of a long-range Coulomb 
interaction $V_\text{coul}(q)$ (green line), a residual
short-range repulsion between quasiparticles $V_\text{res}(q)$ as customary in  Fermi liquid theory (red line), and an attractive interaction $V_\text{attr}(q)$, 
which for simplicity we suppose momentum independent 
($V_\text{attr} = -\lambda$), providing the vertical shift 
highlighted by the orange dashed arrow. The RPA condition 
($q=q_\text{CDW}$) is usually fulfilled away from the 
nesting condition which would occur at the maximum of 
$|\chi^0_{\rho\rho}(q)|$ (corresponding to the kink 
in the black solid line).}
\label{figps}
\end{figure}

This theory does not rely on strong coupling intertwining
between spin and charge degrees of freedom (as it instead occur in the 
stripe phase of 214 cuprates, like, e.g., in Nd- or Eu-codoped 
La$_{2-x}$Sr$_x$CuO$_4$), and therefore is 
an appropriate starting point for the description of CDW in 
YBCO (as well as Bi- and Hg-based cuprates, where the magnitude 
of the CDW wave vector decreases with doping).
A minimal model for this scenario is given by~[\!\!~\cite{cast95,becca,andergassen,sei00}
\begin{eqnarray}
H&=& \sum_{k,\sigma}\varepsilon_k f_{k,\sigma}^\dagger f_{k,\sigma}
-\frac{g}{\sqrt{N}}\sum_{q}(a_q+a_{-q}^\dagger)\rho_q \label{eq:fps} \\
&+&\omega_0 \sum_q a_q^\dagger a_q +\frac{1}{2N}\sum_q [V_\text{coul}(q)+V_\text{res}(q)]\rho_q\rho_{-q}\nonumber
\end{eqnarray}
where operators $f_k^{(\dagger)}$ annihilate (create) quasiparticles that 
correspond to the low-energy sector of a strongly correlated ($U\to \infty$) model.
The dispersion $\varepsilon_k$ contains the nematicity via a parametrization
of the nearest-neighbor hopping $t_{b,a}=t(1\pm \alpha)$ [see Fig. \ref{fscpa}(c)], $\rho_q=\sum_{k,\sigma}
f_{k+q,\sigma}^\dagger f_{k,\sigma}$ is the density operator in momentum space and $N$
denotes the number of lattice sites. The hamiltonian Eq.\,(\ref{eq:fps}) includes a coupling ($\sim g$) to a dispersionless phonon (frequency $\omega_0$) which yields an
effective attractive interaction between quasiparticles $-\lambda\equiv -g^2/\omega_0$. The latter is counteracted by a long-range Coulomb repulsion $V_\text{coul}$ and a residual repulsion between quasiparticles $V_\text{res}(q)$ which is generated by the mapping to the low energy sector, see~\cite{SM} for further details. 

A Fermi liquid which is only subject to a constant (in momentum 
space) attractive interaction $V_\text{tot}(q)=-\lambda$ 
undergoes a transition to a CDW when the 
random-phase-approximation (RPA) condition 
$1/\chi^0_{\rho\rho}(q) - V_\text{tot}(q) = 0$ is met, where 
$\chi^0_{\rho\rho}(q)$ is the charge susceptibility of the 
non-interacting quasiparticle system. Therefore, this instability takes place 
at the dominant nesting wave vector, i.e., the momentum which 
corresponds to the largest charge susceptibility (the kink in 
the black curve in Fig.\,\ref{figps}). Instead, in a strongly 
correlated system, the electron dynamics can be mapped onto an 
effective low energy hamiltonian~\cite{cast95,becca,andergassen,sei00}, see 
Eq. (\ref{eq:fps}), where, according to the Fermi liquid paradigm,
the very large Hubbard repulsion $U$ among electrons is turned into a a residual repulsion between the quasiparticles, $V_\text{res}(q)$ 
(red curve in Fig.\,\ref{figps}), which is maximum at large 
wave vectors, i.e., short distances. Together with an 
attractive contribution $-\lambda$, the total interaction is 
minimum (and negative) at small momenta, and thus fulfills the 
RPA criterion in general at $q = 0$, corresponding to a 
phase separation instability.

On the other hand, the long-range Coulomb repulsion 
$V_\text{coul}(q)$ (green curve in  Fig.\,\ref{figps}) spoils 
the associated zero-momentum instability in the charge sector 
and instead shifts the wave vector of the charge ordering 
transition to finite and in general incommensurate values. 
In a nematic system, all 
these contributions, determining the CDW instability, acquire 
an anisotropic character (analyzed in detail in~\cite{SM}). 
As specified in Eq. (\ref{eq:fps}), we consider an attractive contribution solely due to a momentum independent `Holstein-type' electron-phonon interaction 
$V^\text{ph}(q)\equiv -\lambda$. This parameter has to overcome 
a critical value (orange dashed arrow in Fig.\,\ref{figps}) to 
drive the CDW instability.

\begin{figure}[htbp]
\includegraphics[angle=-0,scale=1]{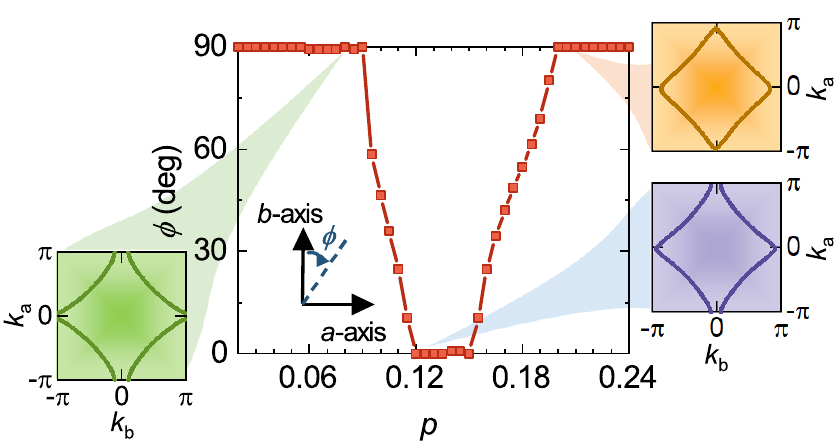}
\caption{Angle (with respect to the $b$-axis) of the CDW 
instability wave vector as a function of doping for a 
Holstein electron-phonon coupling $V^\text{ph}(q)\equiv -\lambda$. The other parameters used in the model, built in the frustrated phase separation scenario, are the following: $a/b=0.98$, $t'/t=-0.17$, $V_\text{coul}/t=0.5$, and the anisotropy parameter  $\alpha=0.015$. In the three insets, the nematic Fermi surfaces obtained using these parameters are shown as a function of doping, in case $p = 0.08$, $p = 0.13$ [similar to the one of Fig.\,\ref{fscpa}(c)] and $p = 0.20$.}
\label{figrpa}
\end{figure}

Over the whole doping range the nematic Fermi surface leads to an enhanced charge susceptibility $\chi^0_{\rho\rho}(q)$ along the $b$-axis. Three different doping regimes can be identified:
\\
- In the low doping regime, the doping dependence of the residual repulsion, $V_\text{res} \sim 1/p$, 
dominates the anisotropy of 
$V_\text{tot}(q)$. Within a standard 
slave-boson approach \cite{SM, cast95, becca, caprara2017},
$V_\text{res}(q)$ turns out to be proportional to 
the quasiparticle kinetic energy.
Since the magnitudes of the hoppings along the $a$- and $b$-axis are
$|t_\text{b}| > |t_\text{a}|$, the repulsion is stronger along
the $b$-axis, so that this term forces the instability to occur along 
the $a$-axis.
\\
- When increasing the doping, the contribution of $V_\text{res}$ to $V_\text{tot}$
decreases  so that the strongly enhanced $\chi^0_{\rho\rho}(q)$ along the $b$-axis, starting at $p\approx 0.12$, leads to a rotation 
of $q_\text{CDW}$  from the $a$- to the $b$-axis of the YBCO unit cell 
(see Fig.\,\ref{figrpa}). Notably, the magnitude of 
$q_\text{CDW}$ does not change across the transition since 
the minimum of $V_\text{tot}(q)$ is only weakly dependent on
the orientation of $q$ ~\cite{SM}. For the considered parameters, the Fermi surface is closed at $(\pi,0)$ while it is still open at $(0,\pi)$ (see violet curve in Fig. \ref{figrpa}). As a consequence, the CDW modulation stays oriented along the $b$-direction in the doping range $0.12\lesssim p \lesssim 0.15$. A strong nesting indeed occurs along the $b$-axis, arising from the small Fermi velocity $v_{kb}$ around $(\pi,0)$ and the  concomitant enhanced density of states. 
\\
-  Finally, at $p\approx 0.15$ the Fermi surface gets also closed at $(0,\pi)$ (while still being nematic, see orange curve in Fig. \ref{figrpa}) and upon increasing doping from $p=0.15$ to $p=0.20$ its topology favors a finite angle between $q_\text{CDW}$ and 
the orthorhombic $b$-axis.~\cite{note_pin}  Therefore the instability vector 
$q_\text{CDW}$ rotates from the $b$- back to the $a$-axis where it remains for $p\gtrsim 0.20$.
In this regime the anisotropic interactions again dominate the orientation of $q_\text{CDW}$ over the
reduced nesting along the $b$-axis.

In summary we have developed a model that accounts for interface effects between HTS thin films and substrate.  In presence of a nanostructured morphology of the substrate surface, uncorrelated atoms will give rise to a strong bond between the films and the substrate. This originates an extra substrate potential which modifies the Fermi surface and induces a CDW instability. The model correctly predicts a nematic Fermi surface and a unidirectional CDW observed in the experiment~\cite{eric20}.  Our finding opens new prospective to tune the ground state of HTS by properly nanopatterning the substrate surface  to induce a substrate potential with well-defined symmetries.

\begin{acknowledgments}
The authors acknowledge support by the Swedish Research Council (VR), 
under the Projects 2018-04658 (F.L.), 2020-04945 (R.A.) and 
2020-05184 (T.B.), by the University of Rome Sapienza, under 
the projects Ateneo 2020 (RM120172A8CC7CC7), Ateneo 
2021 (RM12117A4A7FD11B), Ateneo 
2022 (RM12218162CF9D05), by the Italian Ministero 
dell’Universit\`a e della Ricerca, under the Project PRIN 
2017Z8TS5B and the PNRR MUR project PE0000023-NQSTI (G.M., C.D.C., M.G., S.C.), by the Deutsche 
Forschungsgemeinschaft, under SE 806/20-1 (G.S.).
\end{acknowledgments}


\begin{thebibliography}{99}
\bibitem{note_pin} It should be noted that our model does not include pinning effects which would turn the continuous rotation of the CDW vector again into an abrupt transition from the $b$- to $a$- axis.
\end{thebibliography}
\end{document}